\documentclass[review, 10pt]{elsarticle}

\usepackage{graphicx}
\usepackage{amsmath,amsfonts,amsthm,hyperref}
\usepackage{booktabs}
\usepackage[justification=centering]{caption}
\usepackage{subcaption}
\usepackage{multirow}
\usepackage{xcolor}
\usepackage[flushleft]{threeparttable}
\usepackage{tabulary}
\usepackage{longtable}
\usepackage{lscape}
\usepackage[pass]{geometry}
\usepackage{makecell}
\usepackage{float}

\makeatletter
\def\ps@pprintTitle{%
  \let\@oddhead\@empty
  \let\@evenhead\@empty
  \let\@oddfoot\@empty
  \let\@evenfoot\@oddfoot
}
\makeatother

\biboptions{sort&compress}

\begin{document}

\begin{frontmatter}

\title{Glioma subtype classification from histopathological images using in-domain and out-of-domain transfer learning: An experimental study}

\author[address_1]{Vladimir~Despotovic}
\address[address_1]{Bioinformatics Platform, Data Integration and Analysis Unit, Luxembourg Institute of Health, Strassen, Luxembourg}

\author[address_1]{Sang-Yoon~Kim}

\author[address_2,address_3,address_4,address_5,address_6]{Ann-Christin~Hau}
\address[address_2]{Dr. Senckenberg Institute of Neurooncology, University Hospital Frankfurt, Frankfurt am Main, Germany}
\address[address_3]{Edinger Institute, Institute of Neurology, Goethe University, Frankfurt am Main, Germany}
\address[address_4]{Frankfurt Cancer Institute, Goethe University, Frankfurt am Main, Germany}
\address[address_5]{University Cancer Center Frankfurt, Frankfurt am Main, Germany, University Hospital, Goethe University, Frankfurt am Main, Germany}
\address[address_6]{Laboratoire national de santé, National Center of Pathology, Dudelange, Luxembourg}

\author[address_7]{Aliaksandra~Kakoichankava}
\address[address_7]{Multi-Omics Data Science group, Department of Cancer Research, Luxembourg Institute of Health, Strassen, Luxembourg}

\author[address_8,address_9]{Gilbert~Georg~Klamminger}
\address[address_8]{Luxembourg Centre of Neuropathology, Dudelange, Luxembourg}
\address[address_9]{Klinik für Frauenheilkunde, Geburtshilfe und Reproduktionsmedizin, Saarland University, Homburg, Germany}

\author[address_8,address_10]{Felix~Bruno~Kleine~Borgmann}
\address[address_10]{Department of Cancer Research, Luxembourg Institute of Health, Strassen, Luxembourg}

\author[address_6,address_8]{Katrin~B.~M.~Frauenknecht}

\author[address_6,address_8,address_10,address_11,address_12,address_13]{Michel~Mittelbronn}
\address[address_11]{Luxembourg Centre for Systems Biomedicine, University of Luxembourg, Belval, Luxembourg}
\address[address_12]{Department of Life Sciences and Medicine, University of Luxembourg, Esch-sur-Alzette, Luxembourg}
\address[address_13]{Faculty of Science, Technology and Medicine, University of Luxembourg, Esch-sur-Alzette, Luxembourg}

\author[address_1,address_7]{Petr~V.~Nazarov}

\begin{abstract}
We provide in this paper a comprehensive comparison of various transfer learning strategies and deep learning architectures for computer-aided classification of adult-type diffuse gliomas. We evaluate the generalizability of out-of-domain ImageNet representations for a target domain of histopathological images, and study the impact of in-domain adaptation using self-supervised and multi-task learning approaches for pretraining the models using the medium-to-large scale datasets of histopathological images. A semi-supervised learning approach is furthermore proposed, where the fine-tuned models are utilized to predict the labels of unannotated regions of the whole slide images (WSI). The models are subsequently retrained using the ground-truth labels and weak labels determined in the previous step, providing superior performance in comparison to standard in-domain transfer learning with balanced accuracy of 96.91\% and F1-score 97.07\%, and minimizing the pathologist's efforts for annotation. Finally, we provide a visualization tool working at WSI level which generates heatmaps that highlight tumor areas; thus, providing insights to pathologists concerning the most informative parts of the WSI.
\end{abstract}

\begin{keyword}
digital pathology, whole slide images, glioma, deep learning, transfer learning.
\end{keyword}

\end{frontmatter}


\section{Introduction} 
Histopathology refers to the process of microscopic examination of a surgical specimen or a biopsy to identify the morphological features at the cellular or tissue level that are used either to diagnose a disease, or to estimate its severity and progression. The sampled tissue is fixed in formalin and embedded in paraffin, then sliced into thin sections and affixed onto glass slides \cite{deHaan21}. To distinguish between different cellular components, histological stainings are performed on these sections, the most common being the Hematoxylin and Eosin (H\&E) staining, where hematoxylin stains cell nuclei purple-blue, and eosin stains cytoplasm and connective tissue pink. 

With the advance of technology, digital scanners are used as an alternative to microscopy to scan the samples on glass slides at high resolution and store them as Whole Slide Images (WSI), opening a new epoch of digital pathology \cite{Xu17}. 
In the full resolution and highest magnification, the WSIs often exceed $100000\times100000$ pixels and several gigabytes of storage space. Additionally, several sections are often collected from a single tumor, each investigated with a number of WSIs. Therefore, digital pathology requires the use of advanced image processing techniques for the analysis.
Huge image resolution prevents from using conventional deep learning techniques directly, requiring processing WSIs in a large number of patches or tiles, typically not larger than $512\times512$ pixels.   

In this work, we selected a specific problem in digital pathology, i.e. computer aided diagnostics of adult-type gliomas from WSIs using deep neural networks. 
Diffuse gliomas are the most common type of brain tumors in adults with up to 80\% of primary malignant central nervous system (CNS) tumors \cite{Theeler12}. The last $5^{th}$ edition of the World Health Organization (WHO) classification of CNS tumors released in 2021 proposes a refinement of the classification of adult-type diffuse gliomas based on molecular profiles, largely dependent on isocitrate dehydrogenase 1 or 2 (IDH1/2) mutation status and 1p/19q codeletion status, resulting in 3 primary tumor subtypes: IDH-mutant, 1p/19q codeleted oligodendroglioma; IDH-mutant astrocytoma; and IDH-wildtype glioblastoma \cite{Louis21}. However, the combination of histological and molecular information remains the gold standard for diagnosing and grading CNS tumors \cite{Komori22}. Correct classification of diffuse glioma subtypes is of utmost importance, given that treatment and patient survival largely depend on the tumor subtype, with low-grade gliomas (e.g. oligodendroglioma) reaching 5-year survival rates up to 80\%, and high-grade gliomas (e.g. glioblastoma) below 5\% \cite{Whitfield22}.

Since the diagnosis is based on subjective pathologist assessment, prone to inter- and intra-observer variation, quantitative WSI analysis is essential to improve the reproducibility and diagnostic accuracy, and reduce the workload of pathologists \cite{Laak21, Gurcan09}. 
Given the fact that WSIs are usually labeled only at the slide level (i.e. patient diagnosis is known), and not at the tile level, lack of annotated datasets is a common issue. Large cell heterogeneity and morphological variability represented across different tiles makes the problem even more difficult. Therefore, in circumstances when only small amount of labeled data is available, using transfer learning techniques to pretrain the models in domains where data is easier to acquire and/or annotate is one way to alleviate this problem. 

We investigate two main approaches for training the deep learning models for the automatic classification of histopathological images; i.e. out-of-domain transfer learning and in-domain transfer learning. Out-of-domain transfer learning assumes pretraining the models in the domain of natural images, where large-scale datasets are publicly available and models can be pretrained in a supervised fashion (e.g. ImageNet contains over 1.2 million of images classified into 1000 categories \cite{ILSVRC15}), and either fine-tuning the models using the target WSI dataset annotated at the tile level, or using the pretrained model as a feature extractor. In-domain transfer learning requires pretraining the models on histopathological images that do not necessarily belong to gliomas or even CNS tumors, but can originate from various organs. However, since large-scale histopathological image datasets annotated at the tile level are not available, self-supervised learning approaches are used as an alternative. The idea is to define a domain-specific pretext task that does not require exhaustive annotations by pathologists, where labels are inherent in the source data (e.g. magnification prediction, hematoxylin channel prediction \cite{Koohbanani21}, or cross-stain prediction \cite{Yang21}). 

In this paper, we provide a comprehensive comparison of various in-domain and out-of-domain transfer learning approaches for the classification of adult-type diffuse gliomas from WSIs, and highlight their major advantages and disadvantages. Recent attempts indicate that in-domain transfer learning may outperform out-of-domain pretraining \cite{Chen22, Wang21, Wang2022, Ciga22, Mormont20}; however, to the best of our knowledge, there is no study that attempts to quantify the benefits of in-domain transfer learning in digital pathology in a systematic way. 

To further improve the performance, we propose a two-step semi-supervised learning approach: the pretrained model from the previous step is used to predict the labels of the unannotated tiles in WSIs. The model is subsequently retrained using the ground-truth tiles labeled by the pathologists, augmented with the unlabeled ones annotated by the model; thus, minimizing the efforts for annotation by the trained pathologist. 

Finally, the model at the output of step 2 of semi-supervised learning is used to aggregate the predictions from the tile level to the slide level. The predicted classes are overlaid on the WSI as a heatmap, emphasizing and drawing the pathologist's attention to the most informative areas in the image corresponding to tumor tissue, normal brain tissue or necrosis. The online tool, which can be freely used in preclinical studies, was developed and made available to the community at \url{https://bioinfo.lih.lu/deephisto/}. Furthermore, we present a new dataset for diffuse glioma subtype classification, annotated by experienced pathologists at the tile level, which we use for fine-tuning and evaluation of models. The dataset is available at \url{https://zenodo.org/record/7941080}.

The remainder of the paper is organized as follows. Section~\ref{sec:related_work} provides a discussion on related work. Section~\ref{sec:methods} gives a description of the in-house dataset of adult-type diffuse gliomas used for evaluation, image preprocessing, as well as out-of-domain and in-domain transfer learning techniques, two-step semi-supervised learning, and aggregation of predictions from the tile to the WSI level. Section~\ref{sec:results} provides experimental results, which are further discussed in Section~\ref{sec:discussion}. Study limitations are presented in Section~\ref{sec:limitations}. Finally, Section~\ref{sec:conclusion} gives concluding remarks.

\section{Related work}  
\label{sec:related_work}

Computer-aided image analysis has been recently used for the classification of tumor subtypes \cite{Im21, Lu21} or grading \cite{Mousavi15, Ertosun15} of gliomas, or for survival prediction \cite{Rathore19} from digital histopathological images. Closely related to this task is the prediction of IDH1/2 mutation status, as an important diagnostic and prognostic biomarker in diffuse gliomas \cite{Liu20}. Other methods propose integrating features extracted from histopathological images with molecular features \cite{Pei21, Faust22}, or combining WSIs with Magnetic Resonance Imaging (MRI) \cite{Wang22, Hamidinekoo21}. 

These approaches dominantly use Convolutional Neural Networks (CNN) for extracting features from WSIs, either trained from scratch on the histopathological image dataset of interest \cite{Ertosun15, Liu20, Hamidinekoo21}; using the transfer learning techniques with models pretrained in a domain of natural images \cite{Faust22, Im21}; or handcrafting the image features in some earlier works \cite{Mousavi15}. Another approach is utilized in \cite{Lu21}, where contrastive learning is used to extract discriminative feature representations at the tile level in a self-supervised way, while in the second step, these features are aggregated using a sparse-attention multiple instance learning module for label prediction at the level of WSI.

Although CNNs are still considered to be state-of-the-art models for image classification, a change of paradigm can be recently observed towards the use of attention-based architectures and transformer networks in computer vision, challenging the superiority of CNNs. These can be either hybrid architectures where CNNs are augmented with an attention mechanism to capture long-range dependencies; thus, alleviating the major limitation of CNNs which only operate locally \cite{Bello19}; or pure attention-based transformer models (i.e Vision Transformers (ViT)) \cite{Dosovitskiy21} which can achieve performance comparable to CNNs (or even outperform them on some tasks) without using convolutions, but typically require more training data. Recently introduced Data-efficient image Transformers (DeiT) have shown that with improved data augmentation and regularization strategies, ViTs can be trained with fewer data without any significant changes in architecture \cite{Touvron21}. 

This trend can be also observed in digital pathology. A breast cancer classification model based on color deconvolution and transformer architecture is proposed in \cite{Zhu22}. ViT for tumor detection in sentinel lymph nodes, diffuse large B-cell lymphoma, breast, and lung adenocarcinoma show comparable performance to ResNet50 model \cite{Deininger22}. Hierarchical pyramid ViT architecture leverages a hierarchical structure of morphological features at different image resolutions, starting from $16\times16$ images capturing information at the cell level, $256\times256$ images capturing cell-cell interactions and $4096\times4096$ images representing the cell clusters in tissue micro-environment \cite{Chen22}.

Pretraining the models on datasets of natural images, such as ImageNet, is common in digital pathology \cite{Deininger22}. We refer to this in the remaining text as the out-of-domain pretraining. However, it was shown that when the source and the target domains are not well matched, which certainly is the case when ImageNet pretraining is used in a downstream histopathological domain task, transferability reduces \cite{Yosinski14}. Therefore, efforts were made to use in-domain pretraining using the publicly available annotated datasets, such as Camelyon16, and then transfer the model parameters to another digital pathology task \cite{Khan19, Medela19}. However, Camelyon16 is a relatively small dataset, and the main benefit of transfer learning is observed when models are pretrained on large-scale data. 

In the absence of large-scale datasets of histopathological images annotated at the pixel or tile level, but with the availability of WSIs with known patient diagnoses, self-supervised learning approaches are introduced as an alternative \cite{Chen22, Wang21, Wang2022, Ciga22, Mormont20}. Ciga et al. employ SimCLR, a contrastive self-supervised learning approach, where two augmented versions of the same tile are created and the model parameters are optimized to maximize the similarity between representations of these two versions of the tile (positive examples), at the same time emphasizing dissimilarity to all other tiles in the batch (negative examples) \cite{Ciga22}. 
Positive examples do not necessarily need to come from the same tile. Given the high similarity between certain tiles, instances in different pairs may not always be good candidates for the negative examples, but also for the positive ones. Improved selection of positive examples for contrastive learning by identifying semantically matched instances originating from different tiles increases the visual diversity of positives and enables learning more discriminative feature representations \cite{Wang2022}. 
Wang et al. follow a similar route, but the Bootstrap Your Own Latent (BYOL) strategy was used instead for pretraining, which requires only positive examples to train two networks with identical architectures, but different parameters. The online network predicts the output representation of the target network, which is then iteratively updated as the exponential moving average of the parameters of the online network \cite{Wang21}. Another approach is to use multi-task pretraining by defining a set of classification tasks (both binary and multi-class) for multiple low- and middle-scale histopathological datasets, and training a model that will minimize loss aggregated over all tasks \cite{Mormont20}. 

\section{Material and methods}  
\label{sec:methods}
\subsection{Dataset and study design}
\label{section:dataset}
 
The dataset contains 75 H\&E stained WSIs of 28 adult-type diffuse glioma cases collected at the National Center of Pathology (NCP), Luxembourg National Health Laboratory (Laboratoire national de sant\'{e} - LNS) during the years 2017–2021. 
WSIs were acquired with an IntelliSite Ultra Fast digital slide scanner from Philips containing a 20x/0.75 NA Plan Apo objective with an average slide resolution of 0.25um/pixel.

\begin{table}[thbp]
   \caption{Statistics of the in-house dataset}
   \label{tab:dataset}
   \footnotesize
   \centering
   \begin{tabular}{lcccc}
     \toprule
     \textbf{Participants} & \multicolumn{3}{c}{} \\
     \midrule
     Gender & Male & Female & Total \\
     Number of participants & 18 & 10 & 28 \\
     Age & 57.2 (19) & 49.2 (13.3) & 54.3 (17.6) \\
     \midrule
     IDH-mutant, 1p/19q codeleted (odg*) & 4 & 4 & 8\\
     IDH-mutant (ac**) & 4 & 4 & 8 \\
     IDH-wildtype (gbm***) & 10 & 1 & 11 \\
     Normal brain tissue & 0 & 1 & 1 \\
     \midrule
     \midrule
     \textbf{CNS WHO grade} & 2 & 3 & 4 & NA \\
     \midrule
     IDH-mutant, 1p/19q codeleted (odg*) & 4 & 4 & 0 & 0 \\
     IDH-mutant (ac**) & 2 & 5 & 1 & 0 \\
     IDH-wildtype (gbm***) & 0 & 0 & 11 & 0 \\
     Normal brain tissue & 0 & 0 & 0 & 1 \\
     \bottomrule
   \end{tabular}
    \emph{*odg:} oligodendroglioma;
    \emph{**ac:} astrocytoma; 
    \emph{***gbm:} glioblastoma \\
 \end{table}

The study was approved by the National Research Ethics Committee (Comit\'{e} National d’Ethique de Recherche - CNER; CNER no: REC-LRNO-20110708) and performed according to the General Data Protection Regulation (GDPR) and Declaration of Helsinki. Neuropathological diagnostics of tumor samples (histology, immunohistochemistry, epigenetic or genetic analysis) was performed by a board-certified neuropathologist (MM) according to the $5^{th}$ edition of the WHO classification of CNS \cite{Louis21} and only pseudonymized, region-annotated digital images were exchanged for bio-informatic processing. Three primary tumor subtypes are classified into IDH-mutant, 1p/19q codeleted oligodendroglioma; IDH-mutant astrocytoma; and IDH-wildtype glioblastoma.

\textbf{Tissue Processing, Staining, Imaging and Export}: Patient tissues were routinely fixed directly after operation in 4\% neutral buffered formalin, gradually dehydrated and cleared with an automatic tissue processor, followed by paraffin embedding (formalin-fixed paraffin embedded - FFPE). FFPE tissue blocks were sectioned with a microtome at $3 \mu m$ and $7 \mu m$ thickness, placed on a glass slide, stained with H\&E and scanned with the Philips IntelliSite Ultra Fast scanner. Images were exported as big-tiff images with the following settings: scan factor 20 and quality 80 or 100. Region annotation of WSIs was done by a board-certified pathologist (MM) using the Aperio image scope version 12.3.3 software.

The detailed statistics of the dataset is given in Table~\ref{tab:dataset}. The dataset is dominated by male participants (64.3\%), which correlates with findings that gliomas are 50\% more prevalent in males than in females \cite{Ostrom18}. Grade 4 IDH-wildtype glioblastoma, as the most common malignant primary brain tumor in adults \cite{Desland20}, is also the most prevalent in our dataset (39.3\%). IDH-mutant astrocytoma is represented by 28.6\% of WSIs, out of which 25\% are grade 2, 62.5\% are grade 3 and 12.5\% are grade 4 CNS tumors. IDH-mutant, 1p/19q co-deleted oligodendroglioma is represented by 28.6\% of WSIs, out of which 50\% are grade 2, and 50\% are grade 3 CNS tumors. The brain WSIs of a non-cancer patient were used as normal controls.

\subsection{Data preprocessing}
\label{section:preprocessing}
Given the extremely large resolution of WSIs, they are typically processed in patches or tiles. Regions of interest (ROI) were annotated by marking several rectangular areas in the WSIs, as different levels for normal brain tissue (white and gray matter), necrosis, and the respective tumor entity, e.g. IDH-mutant, 1p/19q codeleted oligodendroglioma; IDH-mutant astrocytoma; and IDH-wildtype glioblastoma. These ROIs are further divided into square 50\% overlapping $512\times512$ tiles, each of them associated with a particular class. 

Image augmentation is applied to all extracted tiles in the training dataset by flipping and rotating by $90^\circ$, $180^\circ$ and $270^\circ$, leading to 8 augmented views of each tile. No augmentation is applied to test data. Basic statistics of the training and test subsets is provided in Table~\ref{tab:train_test}. Note that the training dataset is highly imbalanced with the normal brain class represented with 18 times more images than the necrosis class. This has to be taken into account during the model training and evaluation.

\begin{table}[thbp]
   \caption{Tile level statistics (after image augmentation) of the training and test subsets}
   \label{tab:train_test}
   \footnotesize
   \centering
   \begin{tabular}{lcc}
     \toprule
     \textbf{Class} & \textbf{Training} & \textbf{Test} \\
     \midrule
     \textbf{IDH-mutant, non-codeleted (astrocytoma)} & 30040 & 465 \\
     \textbf{IDH-mutant, 1p/19q codeleted (oligodendroglioma)} & 27072 & 431 \\
     \textbf{IDH-wildtype (glioblastoma)} & 13064 & 241 \\
     \textbf{Normal brain tissue} & 56291 & 2947 \\
     \textbf{Necrosis} & 3112 & 90 \\
     \midrule
     \textbf{Total} & 129579 & 4174 \\
     \bottomrule
   \end{tabular}
 \end{table}

 \subsection{Transfer learning}
\label{section:transfer_learning}

We investigate state-of-the-art transfer learning strategies for computer-aided glioma subtype classification, including supervised out-of-domain transfer learning, where models are pretrained on a large collection of natural images; and in-domain transfer learning, where models are pretrained on medium-to-large-scale publicly available collections of histopathological images. The code and the best performing pretrained models are available at \url{https://git.lih.lu/vdespotovic/deephisto}.

\subsubsection{Out-of-domain transfer learning}
\label{section:out-of-domain}

A variety of CNN-based models (VGG16, VGG19, ResNet18, ResNet50, InceptionV3, MobileNetV2, DenseNet121) pretrained on ImageNet dataset that contains over 1.2 million of natural images was used for out-of-domain transfer learning. VGG networks represent a simple stack of convolutional and max-pooling layers, followed by three fully connected layers on top of them \cite{Simonyan15}. The postfix in the name (16 or 19) refers to the total number of layers. Their major limitation is a large size and a huge number of parameters, making them very slow for both training and inference. 

ResNet architectures are composed of a number of residual blocks, which use the shortcut connections to learn the residuals between input and output. This enables learning more stable networks even with increased depth \cite{He16}. The use of a global average pooling layer instead of fully connected layers leads to a substantial reduction of model parameters in comparison to VGG. The postfix name (18 or 50) refers to the network depth.

The Inception architecture is composed of modules containing parallel concatenations of convolutional layers with filters of different sizes, enabling information extraction at different scales \cite{Szegedy15}. In this way, not only the depth, but also the width of a model can be increased.  

MobileNetV2 is built with 3×3 depthwise separable convolutions, and additionally, it introduces the inverted residual bottleneck layers that allow for memory- and computation-efficient implementation, with significantly fewer parameters in comparison to all previously described networks \cite{Sandler18}.

As opposed to standard CNN architectures where convolutional layers are connected sequentially, in a DenseNet every convolutional layer is connected to all other layers, i.e. the input to a convolutional layer is the concatenation of feature maps from all previous layers \cite{Gao17}. This encourages feature propagation and reuse; thus, significantly reducing the number of parameters.  

Following the latest trends in image classification, we also experiment with Transformer models. In ViTs, the images are divided into tiles, and fed to a standard Transformer network as a sequence of linear embeddings of these patches. To recover the spatial information lost during the flattening, position embeddings are added to patch embeddings \cite{Dosovitskiy21}. We use a base ViT architecture with $16\times16$ patches.

Swin transformers are vision transformers with hierarchical representation that starts from small-sized tiles and gradually increases the size by merging the neighboring tiles. The model uses a shifted window approach which limits the self-attention computation to non-overlapping local windows, but allows connections between neighboring windows, similar to convolutions, preserving at the same time linear computational complexity with respect to image size \cite{Liu21}. We use a tiny version of the Swin transformer, denoted as Swin-T.

Finally, we experiment with the hybrid CNN-Transformer model, where the input sequence for ViT is generated from the feature maps of ResNet50 network \cite{Dosovitskiy21}.

Two strategies for training the models are applied, i.e. training from scratch, and pretraining followed by fine-tuning. In training from scratch, a model with randomly initialized weights is trained directly on the target in-house dataset. In fine-tuning weights and biases of the pretrained model are used to initialize the network, which is then retrained on the target in-house dataset, meaning that all network parameters are updated. 

Details about the pretrained models, including model size, number of model parameters and topological network depth (number of layers in a neural network) are shown in Table~\ref{tab:out-of-domain}. Note that only the layers with trainable weights are considered for calculating the network depth.

\begin{table}[thbp]
   \caption{Deep learning architectures used for out-of-domain transfer learning}
   \label{tab:out-of-domain}
   \footnotesize
   \centering
   \begin{tabular}{lccc}
     \toprule
     \textbf{Model} & \textbf{Size [MB]} & \textbf{Number of parameters} & \textbf{Depth}\\
     \midrule
     VGG16 & 528 & 138.4M & 16 \\
     VGG19 & 548 & 143.7M & 19 \\
     ResNet18 & 45 & 11.7M & 18 \\
     ResNet50 & 97 & 25.6M & 50 \\
     InceptionV3 & 91 & 23.8M & 95 \\
     MobileNetV2 & 14 & 3.5M & 53 \\
     DenseNet121 & 31 & 8M & 121 \\
     ViT-B/16 & 330 & 86.4M & 38 \\
     Swin-T & 108 & 28.3M & 53 \\
     ResNet50-ViT-B/16 & 438 & 114.5M & 102 \\
     \bottomrule
   \end{tabular}
 \end{table}

\subsubsection{In-domain transfer learning}
\label{section:in-domain}

Two strategies are investigated for in-domain transfer learning, i.e. self-supervised and multi-task learning. 

We employ a contrastive self-supervised learning strategy, as given in \cite{Ciga22}, where two stochastically augmented versions of the same tile are created and the model parameters are optimized to maximize the similarity between representations of these two versions of the tile (positive examples). At the same time, dissimilarity to all other tiles in the batch (negative examples) is emphasized using the contrastive NT-Xent loss function. ResNet18 was used as the backbone network to extract the features, and two fully connected layers were added on top of it to transform these features into lower-dimensional image embeddings. The model was pretrained using a collection of 57 histopathological image datasets originating from 22 organs (including The Cancer Genome Atlas Program (TCGA\footnote{https://www.cancer.gov/about-nci/organization/ccg/research/structural-genomics/tcga}) and Clinical Proteomic Tumor Analysis Consortium (CPTAC), as well as multiple publicly available challenge datasets). Different staining methods (H\&E, Wright’s stain, Anti CD3, CD8, Jenner–Giemsa, H-DAB, PAS) and various magnification levels ($10\times$, $20\times$, $40\times$, $100\times$) were used across datasets. For additional information about the datasets, the reader is referred to \cite{Ciga22}.

The second self-supervised learning approach named TransPath \cite{Wang21} employs the Bootstrap Your Own Latent (BYOL) strategy, which does not require negative examples. Two networks with identical architectures, but different weights, i.e. online network and target network, were pretrained using two augmented views of each tile. The online network predicts the output representation of the target network, which is then iteratively updated as the exponential moving average of the parameters of the online network. A hybrid model with ResNet50 as a local feature pre-extractor and ViT that learns global features is used as a backbone network. Models were pretrained using a collection of 32529 WSIs and 2.7 million randomly selected tiles originating from TCGA and Pathology AI Platform (PAIP\footnote{http://www.wisepaip.org/paip/}), covering 25 anatomic sites and 32 cancer subtypes. 

CTransPath \cite{Wang2022} uses a contrastive learning approach built on top of MoCo v3 \cite{Chen21}, but redefines positive examples in the self-supervised learning task, i.e. in addition to an augmented view of the input instance, pseudo-positive semantically relevant examples are selected from a memory bank; thus, increasing the diversity of positives. A hybrid model with a three-layer CNN as a local feature pre-extractor and Swin Transformer that learns global features is used as a backbone network. The same dataset was used for pretraining as in \cite{Wang21}, but containing all tiles, instead of approximately 100 randomly selected tiles from each WSI, leading to a largest available dataset so far composed of more than 15 million of tiles.
 
Another model uses multi-task learning strategy, where the same network architecture is shared across multiple classification tasks that correspond to multiple histopathological image datasets, and a separate classification head is attached to each task \cite{Mormont20}. The classification head is a simple fully connected layer, and the idea behind this choice was to reduce its learning capacity and enforce it to learn generic features  that would be relevant across all tasks. Once the per-task categorical cross-entropy loss is calculated for each task, these losses are aggregated, and the network parameters are optimized to minimize the average total loss. Two CNN architectures are used as a backbone network: ResNet50 and DenseNet121. Models were pretrained on a collection of 22 histopathological datasets for both binary and multi-class classification tasks containing 882800 images in total. Different staining methods were used across datasets, including H\&E, Diff-Quik, May-Grunwald-Giemsa and Masson’s trichrome staining. For additional information about the datasets the reader is referred to \cite{Mormont20}.

In all cases the pretrained network parameters were further transferred for fine-tuning using the in-house dataset. Two fully connected layers are added on top of the backbone network, followed by dropout layers to prevent overfitting. The details about the models used for in-domain transfer learning are provided in Table~\ref{tab:in-domain}.

\begin{table}[thbp]
   \caption{Deep learning architectures used for in-domain transfer learning}
   \label{tab:in-domain}
   \footnotesize
   \centering
   \begin{tabular}{p{2.0cm} p{3.5cm} p{1.5cm} p{1.6cm} p{0.9cm}}
     \toprule
     \textbf{Pretraining} & \textbf{Model} & \textbf{Size [MB]} & \textbf{Parameters} & \textbf{Depth}\\
     \midrule
     \multirow{3}{*}{Self-supervised} & SimCLR (ResNet18) & 45 & 11.7M & 18 \\
     & BYOL (ResNet50+ViT) & 377 & 98.8M & 418\\
     & MoCo v3 (CNN+Swin-T) & 105 & 27.5M & 223\\
     \midrule
     \multirow{2}{*}{Multi-task} & ResNet50 & 97 & 25.6M & 50 \\
     & DenseNet121 & 31 & 8M & 365\\
     \bottomrule
   \end{tabular}
 \end{table}

\subsubsection{Semi-supervised transfer learning}
\label{section:semi-supervised}

We further propose a semi-supervised learning approach, where the fine-tuned models presented in Section~\ref{sec:results_in-domain} are used to predict the labels of the unannotated tiles in WSIs (i.e. the ones not belonging to ROIs). The model is subsequently retrained using the new dataset composed of ground-truth tiles labeled by the pathologists, as well as tiles annotated by the model trained in the previous step. 
 
Since the weak labels are available at the slide level (patient diagnosis), we remove from the new augmented dataset all impossible labels before retraining (e.g. if the particular tile belongs to a slide with a diagnosis of IDH-wildtype (glioblastoma), possible classes are glioblastoma, normal brain tissue or necrosis; tiles labeled with other classes are removed). Furthermore, to prevent using the predictions where the model was insecure, we add only the tiles with the predicted class probability larger or equal 90\%. 

\subsection{Performance metrics}
For the evaluation of models' performances, we use balanced accuracy, precision, recall and F1-score. While in binary classification balanced accuracy is defined as the arithmetic mean of sensitivity and specificity, for multi-class classification problems it equals the macro-average of recall scores per class. For precision, recall and F1-score macro averaging is used, which reduces the multi-class problem to multiple one-vs-all binary predictions, computes the corresponding performance metric per class, and averages the results over all classes. Macro averaging assumes assigning all classes to equal weights, but class imbalance is handled by weighing the loss function instead. Note that in this case the balanced accuracy is defined in the same way as recall.

\section{Results}
\label{sec:results}

\subsection{Out-of-domain transfer learning}
\label{sec:results_out-of-domain}

As a baseline for performance evaluation, we use multiple CNN-based models (VGG16, VGG19, ResNet18, ResNet50, DenseNet121, InceptionV3, MobileNetV2), transformer-based models (ViT-B/16, Swin-T) as well as a hybrid CNN/transformer model (ResNet50-ViT-B/16) trained from scratch on the in-house dataset. Our aim is to evaluate how much pretraining benefits the performance. The models are carefully selected to: 1) test a variety of conceptually different deep learning architectures; and 2) include architectures that are the same (or comparable) to the ones used in out-of-domain transfer learning. 

Before further processing, all images (tiles) are normalized using the in-house dataset statistics, by subtracting the mean and dividing by the standard deviation for each channel. We train models for 10 epochs with a batch size equal to 16. Categorical cross-entropy loss was used as a cost function, and AdamW as an optimizer, with an initial learning rate $10^{-4}$, reduced by a factor of 10 after 5 epochs. Note that AdamW has an adaptive per-parameter learning rate, which is computed using the initial learning rate as an upper limit. Given that dataset is imbalanced, we use categorical cross-entropy loss weighted by the class weights computed as the inverse class frequency of the labels in the training dataset. For estimating the model performance on the test dataset we use the model weights from the best-performing epoch. PyTorch is used to develop and train deep learning models. Models were trained using the NVIDIA Quadro RTX 6000 GPUs. The results for the baseline deep learning models trained from scratch are provided in Table~\ref{tab:out-domain-performance}.

\begin{table}[h]
   \caption{Performance evaluation for glioma subtype classification using out-domain transfer learning (without pretraining and with the models pretrained with ImageNet)}
   \label{tab:out-domain-performance}
   \footnotesize
   \centering
   \begin{tabular}{p{2.0cm} p{2.0cm} p{1.5cm} p{1.2cm} p{1.0cm} p{1.3cm}}
     \toprule
     \textbf{Pretraining} & \textbf{Model} & \textbf{Balanced accuracy} & \textbf{Precision} & \textbf{Recall} & \textbf{F1 score}\\
     \midrule
     \multirow{10}{4cm}{No pretraining} & VGG16 & 85.07 & 83.29 & 85.07 & 82.53 \\
     & VGG19 & 85.00 & 83.75 & 85.00 & 84.34 \\
     & ResNet18 & 88.22 & 79.12 & 88.22 & 82.38 \\
     & ResNet50 & 78.18 & 79.49 & 78.18 & 78.14 \\
     & InceptionV3 & 79.26 & 79.78 & 79.26 & 79.42 \\
     & MobileNetV2 & 84.54 & 85.20 & 84.54 & 84.59 \\
     & DenseNet121 & 85.10 & 85.03 & 85.10 & 84.53 \\
     & ViT-B/16 & 68.25 & 54.62 & 68.25 & 60.09 \\
     & Swin-T & 57.21 & 60.51 & 57.21 & 58.19 \\
     & ResNet50-ViT & 81.02 & 78.34 & 81.02 & 77.22 \\
     \midrule
     \multirow{10}{4cm}{ImageNet} & VGG16 & 94.81 & \textbf{97.04} & 94.81 & \textbf{95.72} \\
     & VGG19 & 93.94 & 94.19 & 93.94 & 93.75 \\
     & ResNet18 & 85.89 & 89.84 & 85.89 & 87.23 \\
     & ResNet50 & 92.42 & 94.65 & 92.42 & 93.50 \\
     & InceptionV3 & 85.70 & 91.02 & 85.70 & 87.25 \\
     & MobileNetV2 & 84.70 & 84.46 & 84.70 & 83.45 \\
     & DenseNet121 & 88.97 & 92.46 & 88.97 & 90.26 \\
     & ViT-B/16 & 92.38 & 93.90 & 92.38 & 92.98 \\
     & Swin-T & 86.22 & 93.82 & 86.22 & 89.16 \\
     & ResNet50-ViT & \textbf{94.87} & 96.63 & \textbf{94.87} & 95.48 \\
     \bottomrule
   \end{tabular}
 \end{table}

To evaluate the impact of the out-of-domain transfer learning, we use the same network architectures, but this time the models were pretrained using the ImageNet dataset. We remove the classification layers and replace them with two fully connected layers with 256 and 128 neurons and ReLU activation, each followed by a dropout equal to 0.5, and an output fully connected layer with softmax activation and 5 neurons corresponding to the number of classes. Note that the same classification head is used for training the models from scratch. To improve the models’ generalization ability to a target dataset containing H\&E stained images, that deviate substantially from ImageNet, we unfreeze all layers, and fine-tune the models with a small learning rate ($5\cdot10^{-6}$), to avoid overfitting. The results for the baseline deep learning models trained from scratch and using the ImageNet pretrained models are provided in Table~\ref{tab:out-domain-performance}.

\subsection{In-domain transfer learning}
\label{sec:results_in-domain}

We use 5 models pretrained on large datasets of histopathological images, either using the self-supervised (SimCLR, BYOL, MoCo v3) or the multi-task learning strategy, as explained in Section~\ref{section:in-domain}. To provide a fair comparison, models were trained using exactly the same setup as in out-of-domain transfer learning (see Section~\ref{sec:results_out-of-domain}). The obtained results are provided in Table~\ref{tab:in-of-domain-performance}.
 
 \begin{table}[thbp]
   \caption{Performance evaluation for glioma subtype classification using in-domain transfer learning}
   \label{tab:in-of-domain-performance}
   \footnotesize
   \centering
   \begin{tabular}{p{1.7cm} p{3.5cm} p{1.5cm} p{1.2cm} p{1.0cm} p{1.3cm}}
     \toprule
     \textbf{Pretraining} & \textbf{Model} & \textbf{Balanced accuracy} & \textbf{Precision} & \textbf{Recall} & \textbf{F1 score}\\
     \midrule
     \multirow{3}{*}{Self-supervised} & SimCLR (ResNet18) & 91.57 & 91.25 & 91.57 & 90.60 \\
     & BYOL (ResNet50+ViT) & \textbf{96.39} & \textbf{95.57} & \textbf{96.39} & \textbf{95.81} \\
     & MoCo v3 (CNN+Swin-T) & 93.45 & 94.65 & 93.45 & 93.45 \\
     \midrule
     \multirow{2}{*}{Multi-task} & ResNet50 & 93.15 & 92.86 & 93.15 & 92.54 \\
     & DenseNet121 & 94.31 & 89.66 & 94.31 & 91.43\\
     \bottomrule
   \end{tabular}
 \end{table}
 
\subsection{Semi-supervised in-domain transfer learning}
\label{sec:results_semi-supervised}

\begin{figure}[h!]
 \centering
    \begin{center}
        \includegraphics[width=8.5cm]{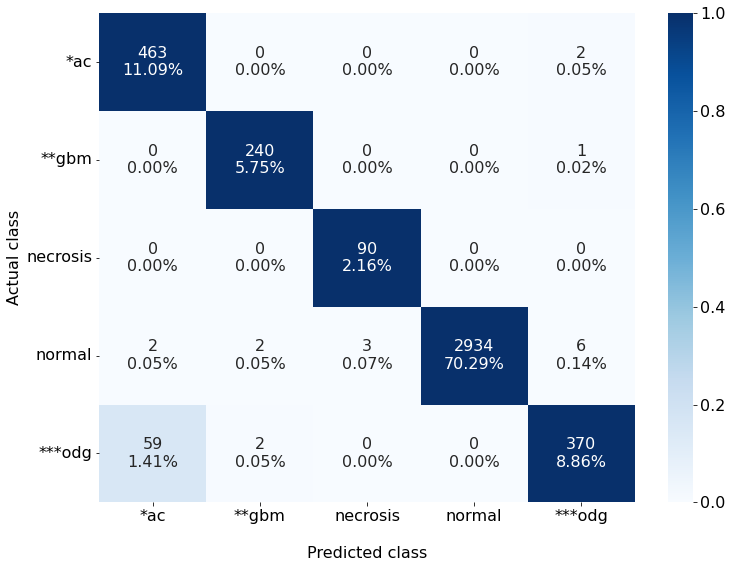}
    \end{center}
    \caption{Confusion matrix for the best performing in-domain transfer learning model (semi-supervised ResNet50+ViT).}
    \footnotesize
    \emph{*ac:} IDH-mutant (astrocytoma); 
    \emph{**gbm:} IDH-wildtype (glioblastoma); \\
    \emph{***odg:} IDH-mutant, 1p/19q codeleted (oligodendroglioma) \\
    \emph{Percentages correspond to the fractions of total number of images in the test dataset}
    \label{fig:confmat}
\end{figure}

\begin{figure}[h!]
 \centering
    \begin{center}
        \includegraphics[width=11cm]{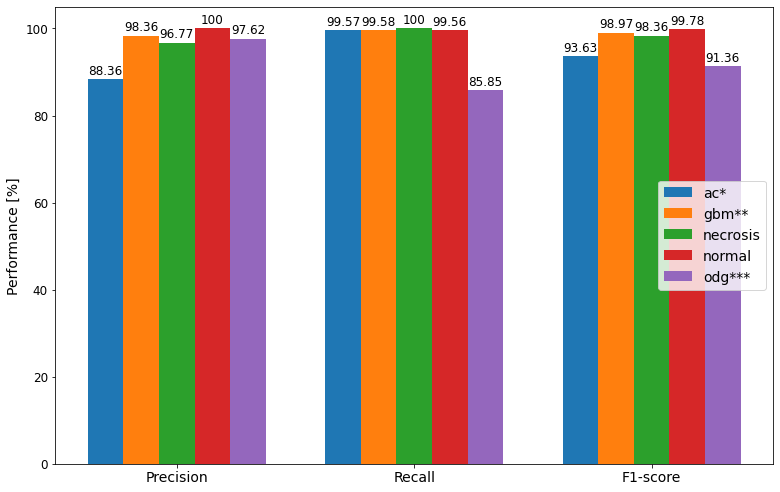}
    \end{center}
    \caption{Per class performance for the best performing in-domain transfer learning model (semi-supervised ResNet50+ViT).}
    \scriptsize
    \emph{*ac:} IDH-mutant (astrocytoma); 
    \emph{**gbm:} IDH-wildtype (glioblastoma); \\
    \emph{***odg:} IDH-mutant, 1p/19q codeleted (oligodendroglioma) \\
    \label{fig:barplot}
\end{figure}

Given the improved performance of the models pretrained in-domain in comparison to out-of-domain transfer learning, we decide to run a two-step semi-supervised learning approach only for the models pretrained in-domain. The summary of results for the in-domain semi-supervised learning is provided in Table~\ref{tab:in-of-domain-performance_semi}, whereas confusion matrix and per class performance of the best model (ResNet50 + ViT-B/16) are provided in Figure~\ref{fig:confmat} and Figure~\ref{fig:barplot}, respectively.

We also provide the size of the augmented training datasets after the first step of semi-supervised learning in Table~\ref{tab:in-of-domain-datasize_semi} for all in-domain models. Note that the training data size was increased from 2.2 times for SimCLR up to 3.2 times for MoCo v3 in comparison to size of the ground-truth dataset. 
 
  \begin{table}[h]
   \caption{Performance evaluation for glioma subtype classification using in-domain transfer learning in a semi-supervised learning scenario}
   \label{tab:in-of-domain-performance_semi}
   \footnotesize
   \centering
   \begin{tabular}{p{1.7cm} p{3.5cm} p{1.5cm} p{1.2cm} p{1.0cm} p{1.3cm}}
     \toprule
     \textbf{Pretraining} & \textbf{Model} & \textbf{Balanced accuracy} & \textbf{Precision} & \textbf{Recall} & \textbf{F1 score}\\
     \midrule
     \multirow{3}{*}{Self-supervised} & SimCLR (ResNet18) & 96.55 & 95.27 & 96.55 & 95.62 \\
     & BYOL (ResNet50+ViT) & \textbf{96.91} & 96.22 & \textbf{96.91} & 96.42 \\
     & MoCoV3 (CNN+Swin-T) & 96.56 & 97.03 & 96.56 & 96.58 \\
     \midrule
     \multirow{2}{*}{Multi-task} & ResNet50 & 96.48 & \textbf{97.94} & 96.48 & \textbf{97.07} \\
     & DenseNet121 & 96.63 & 93.01 & 96.63 & 94.57 \\
     \bottomrule
   \end{tabular}
 \end{table}

 \begin{table}[H]
   \caption{Size of the training dataset after augmentation with semi-supervised labels}
   \label{tab:in-of-domain-datasize_semi}
   \footnotesize
   \centering
   \begin{tabular}{llc}
     \toprule
     \textbf{Type of pretraining} & \textbf{Model} & \textbf{Dataset size} \\
     \midrule
     \multirow{3}{*}{Self-supervised learning} & SimCLR (ResNet18) & 284606 \\
     & BYOL (ResNet50+ViT) & 416826 \\
     & MoCo v3 (CNN+Swin-T) & 402347 \\
     \midrule
     \multirow{2}{*}{Multi-task learning} & ResNet50 & 377776\\
     & DenseNet121 & 339046 \\
     \bottomrule
   \end{tabular}
\end{table}
 
\subsection{Quantitative localization of diffuse gliomas in Whole Slide Images}
\label{sec:results_localization}

\begin{figure}[H]
\centering
\begin{subfigure}[b]{0.62\textwidth}
   \includegraphics[width=1\linewidth]{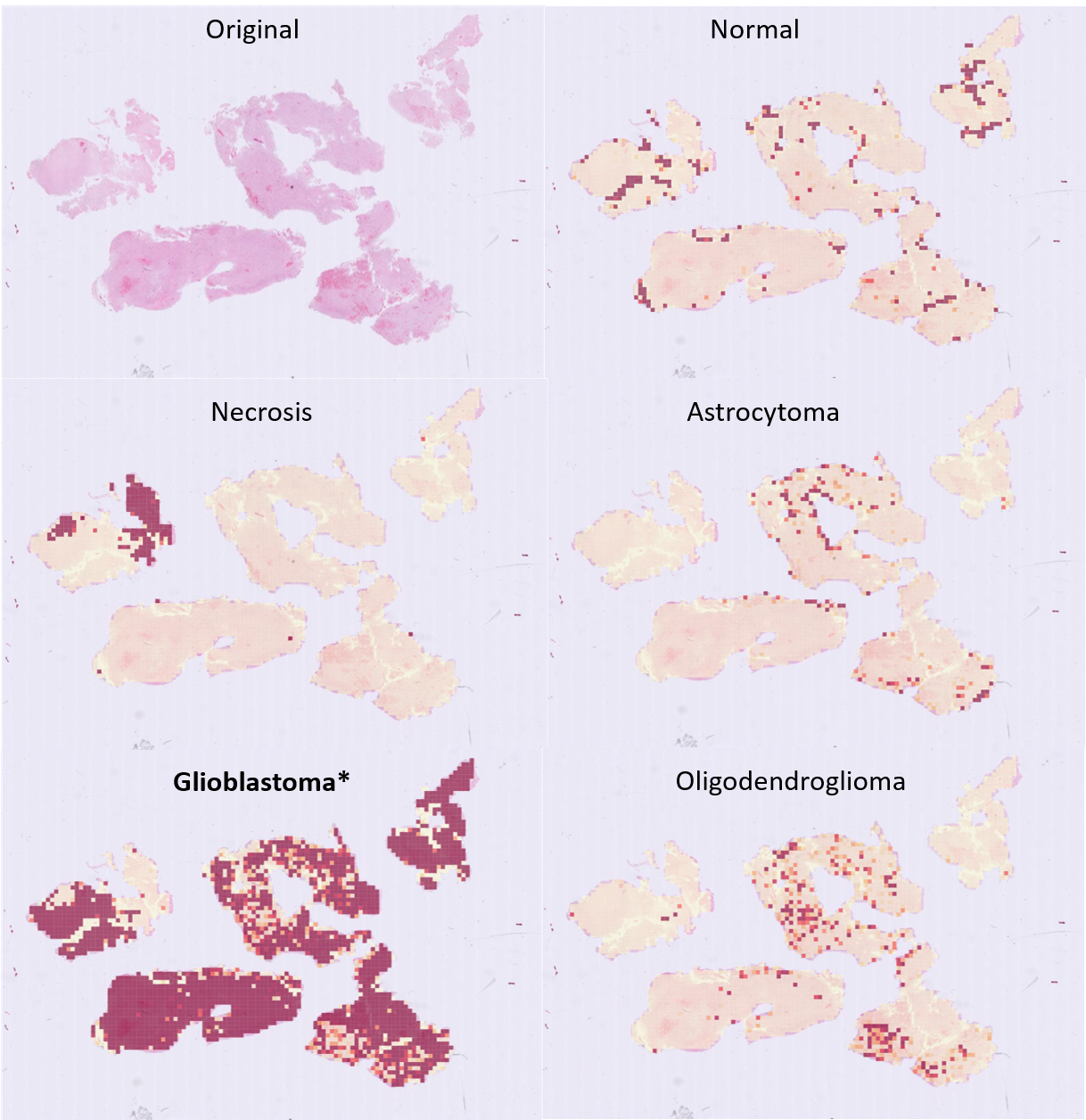}
   \caption{}
   \label{fig:heatmap_gbm} 
\end{subfigure}
\begin{subfigure}[b]{0.62\textwidth}
   \includegraphics[width=1\linewidth]{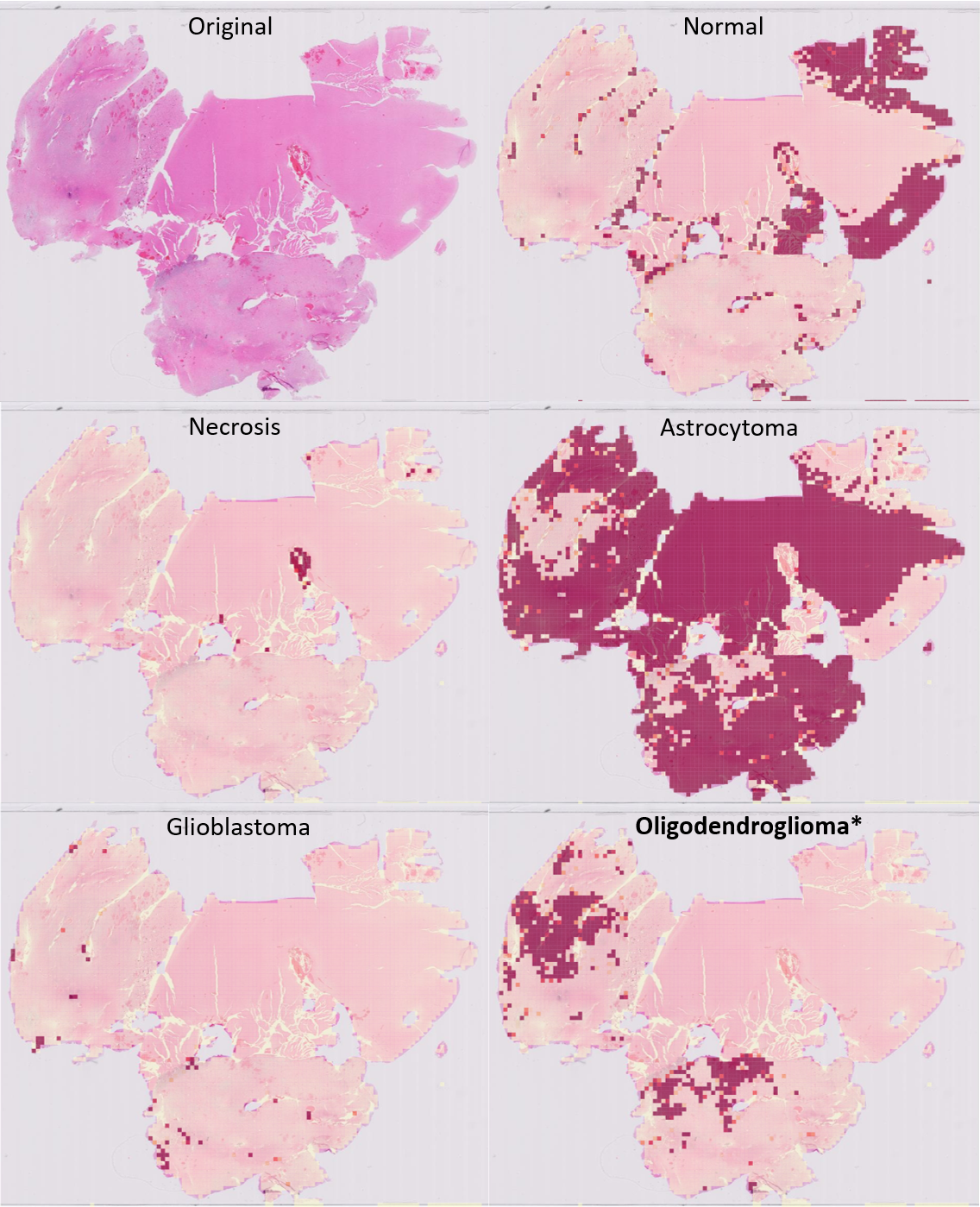}
   \caption{}
   \label{fig:heatmap_oligo}
\end{subfigure}
\caption{WSI level prediction (semi-supervised ResNet50+ViT) \\
\footnotesize
(a) IDH-wildtype (glioblastoma); (b) IDH-mutant, 1p/19q codeleted (oligodendroglioma). \\ 
*Correct class}
\label{fig:heatmap}
\end{figure}

To aggregate the predictions from the tile level to the WSI level, we overlay the predicted confidence scores for each tile on the WSI as a heatmap, where the red color corresponds to the classes, i.e. 3 diffuse glioma subtypes, normal brain tissue and necrosis, as shown in Figure~\ref{fig:heatmap}. The idea is to draw the pathologist’s attention to the most informative areas in the WSI corresponding to the tumor tissue.

Figure~\ref{fig:heatmap_gbm} shows an example of the correct classification of the IDH-wildtype (glioblastoma) tumor using the hybrid semi-supervised ResNet50+ViT-B/16 model, whereas Figure~\ref{fig:heatmap_oligo} presents the case when the model is inconclusive between the IDH-mutant, 1p/19q codeleted (oligodendroglioma) and IDH-mutant (astrocytoma). 

We furthermore provide an online tool named DeepHisto freely available to the community (\url{https://bioinfo.lih.lu/deephisto/}), where the best performing model (hybrid semi-supervised ResNet50+ViT-B/16 model) can be used for localization of diffuse gliomas in WSIs. 

\section{Discussion}
\label{sec:discussion}

Out-of-domain transfer learning using the deep neural networks pretrained on ImageNet dataset as "off-the-shelf' feature extractor has become dominant in digital pathology, successfully applied for prediction of tissue types, molecular features, and clinical outcomes \cite{Li22}. Despite the fact that the models are pretrained in an entirely different domain of natural images, it was shown that models provide satisfactory performance, outperforming the models trained from scratch with limited histopathological image datasets \cite{Kieffer17}. Another option is to use a model pretrained on ImageNet for weight initialization, and unfreeze the network layers for fine-tuning using the smaller histopathological image dataset, which typically leads to improved performance and faster convergence \cite{Mormont18, Otalora22}. Our results in Table~\ref{tab:out-domain-performance} confirm these findings, showing substantial performance improvement for both CNN-based and transformer models, that goes up to 32\% for Swin-T measured by F1-score. The best-performing model is ResNet50-ViT-B/16 with a balanced accuracy of 94.87\%, and F1-score of 95.48\%. Comparing these results to Table~\ref{tab:out-of-domain} one can observe that smaller models with fewer parameters, such as MobileNet or DenseNet, benefit less from fine-tuning. We furthermore observe that transformer-based models (ViT-B/16, Swin-T) reach modest performance in comparison to CNN-based models when trained from scratch, due to the fact that transformers typically require data on a larger scale for pretraining to surpass the lack of inductive biases, such as locality or translational equivariance, which are embedded in CNNs \cite{Dosovitskiy21}. However, with ImageNet pretraining their performance is comparable to CNN-based models.

Further experiments are done with models entirely trained on a large collection of histopathological image datasets (in-domain transfer learning), either using a semi-supervised approach, or multi-task learning paradigm, as explained in Section~\ref{section:in-domain}. Recent attempts indicate that in-domain transfer learning outperforms out-of-domain transfer learning when sufficiently enough domain-relevant data is available for pretraining \cite{Wang21, Wang2022, Ciga22, Mormont20}. In order to test this hypothesis in a systematic manner, we use all models pretrained on large-scale histopathological data that are available up-to-date (to the best of our knowledge), and fine-tune them under identical conditions on the in-house dataset for the task of prediction of diffuse glioma subtypes. The results presented in Table~\ref{tab:in-of-domain-performance} show improved performance in comparison to models pretrained on ImageNet (Table~\ref{tab:out-domain-performance}), but to a lower extent than one would expect, given the large mismatch between the domains of natural and histopathological images. Even for the models trained using the same network architectures, there is only a slight improvement of 0.4\% in F1-score for the ResNet50-ViT-B/16 trained with BYOL, around 1\% for the DenseNet121 model (trained with multi-task learning), and around 3\% for remaining self-supervised learning models, where interestingly for the ResNet50 model (trained with multi-task learning) the performance even slightly drops. This suggests that the learned image features are mostly domain-invariant, requiring only gentle fine-tuning in the target domain with a small learning rate for a limited number of epochs. Relative improvement for self-supervised learning models is, in general, higher than for multi-task learning, but one should also take into consideration that they are pretrained on substantially larger datasets. 

Additionally, we want to investigate whether a two-step semi-supervised training, which would augment the initial target dataset with the weakly labeled tiles learned by the model, can boost the model performance even further. Semi-supervised learning improves the performance for all investigated models (see Table~\ref{tab:in-of-domain-performance_semi}), with improvements ranging from 0.6\% for a hybrid ResNet50-ViT-B/16 model (pretrained with BYOL), up to 5\% for ResNet18 (pretrained with SimCLR). The performance of the best model reaches balanced accuracy of 96.91\% and F1 score of 97.07\%. 

In order to analyze performance for individual classes we plot in Figure~\ref{fig:confmat} the confusion matrix of the best performing model (ResNet50-ViT-B/16 model), concluding that training with the categorical cross-entropy loss function weighted by the class weights has solved the imbalanced data issue, with almost perfect classification for all classes (including the minority classes, such as necrosis), except the IDH-mutant, 1p/19q codeleted (oligodendroglioma) class which is in 16\% of cases misclassified as IDH-mutant (astrocytoma). However, due to their morphological similarity, this represents a challenge even for trained pathologists, with commonly confounded diagnoses and large intraobserver variability \cite{Hsu22}. 
Analysing per class performance in Figure~\ref{fig:barplot} similar conclusion can be drawn. F1-score is almost perfect for all classes, except astrocytoma and oligodendroglioma. Precision for astrocytoma is approximately 10\% lower than for the remaining classes, due to an increased number of false positives (swapping with oligodendroglioma), whereas recall is 15\% decreased for oligodendroglioma due to an increased number of false negatives (swapping with astrocytoma). Another interesting observation can be made in Figure~\ref{fig:confmat}: if the model would be used as a binary classifier for tumor detection (class "no tumor" corresponding to normal tissue/necrosis, and "tumor" corresponding to all tumor subtypes), the probability of a false alarm (detecting cancer in normal tissue) is only 0.33\%, whereas the probability of misclassification of cancer is equal to 0.

Finally, we generate a slide-level prediction by overlaying the predicted confidence scores for each tile on the WSI as a heatmap, as presented in Figure~\ref{fig:heatmap}. 
Figure~\ref{fig:heatmap_gbm} shows an example of the correct classification, where most of the tumor tissue is correctly classified as IDH-wildtype (glioblastoma) using the hybrid semi-supervised ResNet50+ViT-B/16 model, with only minor incorrectly classified areas of IDH-mutant (astrocytoma) and IDH-mutant 1p/19q codeleted (oligodendroglioma). Necrotic patterns as a hallmark feature of glioblastoma are also correctly recognized, which is of particular diagnostic interest since it correlates with tumor aggressiveness and poor prognosis \cite{Yee20}. However, there are cases where the model is inconclusive and misclassifies areas of IDH-mutant 1p/19q codeleted (oligodendroglioma) as IDH-mutant (astrocytoma), as a result of their morphological similarity (see Figure~\ref{fig:heatmap_oligo}).

\section{Study limitations}
\label{sec:limitations}

While the performance results obtained in this study are very high, confirming the potential of using AI in digital pathology, it should be noted that the dataset used for model evaluation is relatively small and may suffer from the limited diversity in terms of number of patients and different scanners being used. Unfortunately, there is a lack of publicly available datasets annotated at the tile level. 
On the other hand, more diverse WSIs acquired using different scanners from different sources, may also introduce batch effects, where deep learning model tends to learn slide origin, scanner type, or slide preparation (e.g. differences in sample processing and staining), rather than predicting the outcome of interest \cite{Bankhead22}.
However, the primary objective of this study was not to maximise the performance, but to provide a rigorous and fair comparison of various transfer learning strategies for glioma subtype classification from histopathological images. Therefore, limiting potential batch effects might be even preferred in such setup.

\section{Conclusion}
\label{sec:conclusion}

The paper investigates different transfer learning strategies in digital pathology and conducts comprehensive experiments for diffuse glioma subtype classification, comparing in-domain transfer learning, where models are pretrained on medium-to-large-scale publicly available collections of histopathological images in a self-supervised or a multi-task learning manner; with supervised out-of-domain transfer learning, where models are pretrained on a large collection of natural images (ImageNet). Although in-domain transfer learning provides certain performance improvement, we found that concerns regarding the generalizability of ImageNet representations for the domain of histopathological images are not entirely justified, showing that after fine-tuning on the target task, reasonable performance can be obtained in the absence of large scale in-domain data for pretraining the models. 

We further assessed in-domain models in a semi-supervised learning scenario, by augmenting the target dataset composed of ROIs annotated by pathologists with areas of WSIs outside ROIs labeled by the model. Semi-supervised learning achieves substantial performance improvement of up to 5\%, thus minimizing the pathologists' efforts for annotation. However, this comes at the expense of increased complexity, requiring a two-step learning approach.  

Finally, we provide visualizations at the WSI level, by generating heatmaps that localize and highlight the tumor areas, therefore drawing pathologist’s attention to the most informative areas of the WSI.

\section{Acknowledgements}
This work was supported by the Luxembourg National Research Fund (FNR) (C21/BM/15739125/DIOMEDES to P.V.N. and PEARL P16/BM/11192868 grant to M.M.); A-C.H was funded by the Mildred-Scheel Career Center Frankfurt (Deutsche Krebshilfe).

\bibliography{manuscript}

\end{document}